\documentclass[a4paper,11pt]{article}

\input{xy}
\xyoption{all}
\xyoption{poly}
\usepackage[all]{xy}
\usepackage{amsmath,amsthm,amsfonts,amssymb,stmaryrd,fancyhdr,mathtools}
\usepackage{latexsym}
\usepackage{color}
\usepackage{graphicx}
\usepackage{float}
\usepackage{fullpage}
\usepackage{mdframed}
\usepackage[font=small,labelfont=small]{caption}
\usepackage[T1]{fontenc}
\usepackage[utf8]{inputenc}
\usepackage{authblk}
\usepackage{hyperref}
\usepackage{cite}
\usepackage{fancyhdr}
\numberwithin{equation}{section}

\allowdisplaybreaks

\fancypagestyle{plain}{%
    \fancyhead[R]{MPP-2019-261}
    
}

\title{Winding modes and the reflection symmetry in AdS$_3$}
\date{December 20, 2019}
\author[1,2,3]{Sergio~M.~Iguri\thanks{siguri@iafe.uba.ar}}
\author[4]{Nicol\'as~Kovensky\thanks{n.kovensky@soton.ac.uk}}
\author[5]{Leila~Maestri\thanks{maestri@mpp.mpg.de}}
\author[1,2]{Lucas~Mart\'in\thanks{lucasmartinar@iafe.uba.ar}}
\affil[1]{Universidad de Buenos Aires, Facultad de Ciencias Exactas y
Naturales. Ciudad Universitaria, 1428 Buenos Aires, Argentina.}
\affil[2]{CONICET-Universidad de Buenos Aires, Instituto de Astronomía
y Física del Espacio (IAFE). C.~C.~67, Suc.~28, 1428 Buenos Aires, Argentina.}
\affil[3]{Universidad Abierta Interamericana, Facultad de Arquitectura. 1428 Buenos Aires, Argentina}
\affil[4]{ University of Southampton, Mathematical Sciences and STAG Research Centre. 
\newline
SO17 1BJ Southampton, United Kingdom.}
\affil[5]{Max-Planck-Institut f\"ur Physik, Werner-Heisenberg-Institut. D-80805 Munich, Germany.}

\begin{document}

\maketitle

\begin{abstract}

We compute the contact term of the two-point function for the SL($2,\mathbb R$)-WZNW model in the winding sector. After reviewing some generalities of the model and its Euclidean counterpart, we discuss the reflection symmetry for the spectral flowed sector. We define the corresponding intertwining operator and use it to find the explicit form of the contact term of the propagator for a vertex carrying an arbitrary amount of spectral flow. Finally, we enhance the already known integral expression of a singly flowed vertex in order to reproduce such contact term directly in the space-time picture.

\end{abstract}


\bigskip

\section{Introduction}

The study of strings propagating in AdS$_3$ has generated many interesting insights, ranging from string theory itself, integrability and AdS$_3$ quantum gravity \cite{Seiberg:1990eb,Witten:1988hc,Witten:1991yr,Mukhi:1993zb,Ghoshal:1995wm,Giveon:1999px,Giveon:1999tq,Banados:1992wn,Beisert:2010jr} to Liouville theory and condensed-matter physics \cite{Zirnbauer:1996zz,Zirnbauer:1999ua,Bhaseen:1999nm,Kogan:1999hz}. When the background is purely of the NS-NS type, the worldsheet theory is given by the SL($2,\mathbb R$)-WZNW model, a theory that has been intensively studied since Maldacena's conjecture \cite{Maldacena:1997re,Witten:1998qj} becoming, so far, one of the few known theoretical schemes in which the AdS/CFT correspondence may be explored beyond the supergravity approximation. Furthermore, computations on the AdS$_3$ side can be compared with the two-dimensional CFT description, where powerful techniques are also available, thus giving, in principle, the possibility of constructing dual pairs where both sides can be exactly solved.   

Over the last couple of years there has been a renewed interest in this type of AdS/CFT scenario. These efforts are mainly concentrated on the so-called $k=1$ sector, \textit{i.e.} the case with minimal AdS flux. The propagation of strings in space-times of the form AdS$_3\times S^3 \times {\cal{M}}_4$ was considered, including, for example, ${\cal{M}}_4 = T^4$, K3 and $S^3 \times S^1$, and it was argued that the corresponding field theory duals were given by deformations of symmetric orbifold CFT's \cite{Gaberdiel:2017oqg,Eberhardt:2017pty,Eberhardt:2018ouy,Eberhardt:2019niq}. 
Part of these considerations remain valid for the $k>1$ case, albeit now with symmetric orbifold theories containing additional Liouville factors \cite{Eberhardt:2019qcl}. 

Unlike the case of a  WZNW model having a compact underlying symmetry,  when the target space is the universal cover of the SL($2,\mathbb R$) group manifold, the spectrum standardly constructed upon irreducible representations of the zero-mode algebra must be enhanced to avoid a coupling independent restriction on the masses of physical states and to give account of long string configurations, {\em i.e.} finite energy states classically corresponding to strings stretched close to the boundary of AdS$_3$ \cite{Maldacena:1998bw,Maldacena:1998uz,Seiberg:1999xz}.

The key ingredient for generating the full physical spectrum is the {\em spectral flow} \cite{Maldacena:2000hw,Maldacena:2000kv,Maldacena:2001km,Israel:2003ry}, a family of automorphisms of the current algebra, labelled by an integer number $\omega$, the so-called spectral flow number or charge, which, in some cases, can be recognized as the amount of winding of a long string along the angular direction of AdS$_3$. For a rational WZNW model, spectral flow trivializes, as it relates standard representations, mapping primary states of one into the current algebra descendants of another. In the SL($2,\mathbb R$)-WZNW model, a Lorentzian non-rational CFT, modules with different spectral flow charges turn out to be generically non-equivalent, spectral flow automorphisms thus defining new representations. Operators with non-trivial spectral flow describing the \textit{winding sector} of the theory were shown to play an important role in the models described above.

Correlators involving only spectrally unflowed vertex operators are obtained from the correlation functions in the H$_3^+$-WZNW model \cite{Teschner:1997ft,Teschner:1999ug}, the Euclidean counterpart of the SL($2,\mathbb R$)-WZNW model, by analytic continuation. However, more care must be taken when dealing with amplitudes involving spectral flowed insertions. There are two known strategies for computing these correlation functions, both exploiting the singular properties of the so-called spectral flow operator.

Regarding the first one, the computation is performed in the original space-time picture. Roughly speaking, every vertex operator associated with a state carrying a single unit of spectral flow is expressed as an unflowed vertex convoluted with a spectral flow operator, the corresponding integral being understood to hold while inside a correlator. This integral definition was introduced and subsequently used for determining the regular term of the propagator of two $\omega=1$ states and the three-point function involving two spectrally unflowed states and one vertex with unit spectral flow in \cite{Maldacena:2001km}. 
The main restraint of the procedure comes from the fact that the referred integral definition of a spectral flowed vertex exists, so far, only for operators with a single unit of spectral flow. The generalization for an arbitrary charge of spectral flow is still lacking.

The second strategy is the so-called FZZ procedure. It was firstly developed in \cite{Fateev} based on parafermionic operators and the properties of their correlation functions. According to it, starting with a regular unflowed correlator, a spectral flow operator is inserted for each unit of spectral flow carried by each vertex. After Mellin-transforming the amplitude thus obtained to the basis in which the Cartan generator of SL($2,\mathbb R$) is diagonal, also called the $m$-basis for short, the dependence on the ``unphysical'' insertion points is removed and the worldsheet dependence, properly adjusted. The computation concludes after transforming back to space-time picture or $x$-basis. No constraint on the value of $\omega$ is imposed.


Transforming correlation functions in the $m$-basis back to the space-time picture is not a simple task. In some cases in which the affine symmetry dictates the functional dependence of a correlator on the space-time coordinates, as for the regular term of the propagator or some three-point functions, no inverse Mellin transformation is needed and the FZZ recipe can be easily carried out. However, knowing the dependence of a correlator on the space-time coordinates may not be enough for the FZZ procedure to be fully completed.

In this paper, we compute the contact term of the propagator in a general setting. Such term is expected to exist in a spectral flowed frame as it already exists for unflowed vertex operators, its occurrence being necessary in order to properly normalize spectral flowed states in the $x$-basis. Our strategy for determining this singular term relies on a generalization of the reflection symmetry in sectors with $\omega\ne 0$, since a single reflection in a two-point function should switch the corresponding regular and contact terms.

Reflection in the Euclidean model is a Weyl-type symmetry expressing the unitary equivalence of certain irreducible representations of SL($2,\mathbb C$) and the affine modules constructed upon them. In the Lorentzian case, in the $m$-basis, this interpretation is retained for the family of continuous series, since the generalized reflection symmetry intertwines between the corresponding affine modules and their spectrally flowed images. In the space-time picture as well as for discrete representations in $m$-basis and their related modules, reflection is recognized as a $\mathbb Z_2$ symmetry relating microscopic states in the sense of \cite{Teschner:1999ug}.

The final expression we get for the two-point function deserves a couple of comments. Of course, it reduces to the known propagator in the unflowed limit, {\em i.e.} $\omega \rightarrow 0$. On the other hand, its dependence on the worldsheet coordinates consistently shows that, although they retain some conformal symmetries, the vertex operators in the space-time picture are not even Virasoro quasi-primary fields. Finally, the form of the overall constant in the propagator could be relevant for describing transport properties of conformal field theories associated by duality with string physics in AdS$_3$, a subject that  has recently aroused interest \cite{Polchinski:2012nh,Hartnoll:2016apf}. Indeed, this constant admits a factorized form that is well suited for a proper normalization of spectral flowed vertex operators in the space-time scenario. This normalization become singular for some configurations suggesting that it could have an impact while studying singularities of the propagator and their interpretation in dual models. 

The paper is organized as follows. In sections \ref{sect2} and \ref{sect3} we review some generalities of the H$_3^+$-WZNW model and its Lorentzian counterpart. We stress that by the latter we understand the WZNW model whose target space is the universal cover of the SL$(2,\mathbb R)$ group manifold. After introducing the spectra of both theories, we discuss the reflection symmetry in the Euclidean model and its emergence in the SL$(2,\mathbb R)$-WZNW model case, including the winding sector. In section \ref{sect4}, we introduce the vertex operators associated with spectrally flowed states and we review the computation of the regular term of the propagator in the space-time picture as done in \cite{Maldacena:2001km}. We also clarify some aspects regarding the definition of spectral flowed vertex operators in the $x$-basis. In section \ref{sect5}, we generalize the expression of the reflection symmetry for a sector with $\omega \ne 0$, and use it to explicitly compute the contact term of the propagator in section \ref{sect6}. In section \ref{sect7}, we discuss the impact of the reflection symmetry on the integral definition of a vertex with a single spectral flow charge. Finally, in section \ref{disc}, we present our conclusions.

\section{The H$_3^+$-WZNW model and the reflection symmetry}
\label{sect2}

Although the WZNW model describing string propagation in AdS$_3$ is the one based on the universal covering group of SL$(2,\mathbb R)$, some aspects of the theory can be read off more easily from its Euclidean counterpart, namely, the H$_3^+$-WZNW model. In this section, we review some basics of the H$_3^+$-WZNW model, as many of the formulas obtained in this context remain valid for the unflowed sector of the SL$(2,\mathbb R)$-WZNW model once the dependence on the space-time momentum is analytically continued. Most of these formulas will be properly generalized for sectors with non trivial spectral flow number in later sections. We shall follow \cite{Teschner:1997ft} closely.

The spectrum $\mathcal V_{\mbox{\scriptsize H}_3^+}$ of the H$_3^+$-WZNW model carries a representation of two commuting isomorphic $\mathfrak{sl}(2,\mathbb C)$ current algebras generated by the modes $J^a_n$ and $\bar J^a_n$, with $a=+,0,-$ and $n\in\mathbb Z$. The holomorphic modes satisfy 
\begin{eqnarray}
&&\left[J^0_n,J^0_m\right]=-\frac{1}{2}k  n \delta_{n+m,0},\nonumber\\
&&\left[J^0_n,J^{\pm}_m\right]=\pm J^{\pm}_{n+m},\nonumber\\
&&\left[J^-_n,J^+_m\right]=2J^{0}_{n+m}+k n \delta_{n+m,0},\nonumber
\end{eqnarray}
where $k$ denotes the level of the current algebra. Identical relations hold for the antiholomorphic generators as well.

As usual, there are two commuting Virasoro algebras in the universal enveloping algebra of the current algebra with generators $L_n$ and $\bar L_n$, with $n\in\mathbb Z$, defined according to the Sugawara construction, namely,
\begin{equation}
L_n=\frac{1}{2(k-2)}\sum_{k\in\mathbb Z}:J^+_{-k}J^-_{n+k}+J^-_{-k}J^+_{n+k}-2J^0_{-k}J^0_{n+k}:,\nonumber
\end{equation}
where the normal ordering is defined as
\begin{equation}
:J^a_nJ^b_m:=\left\{\begin{array}{ll}
J^a_nJ^b_m&\mbox{if}~n<m,\\
\frac{1}{2}\left(J^a_nJ^b_n+J^b_nJ^a_n\right)&\mbox{if}~n=m,\\
J^b_mJ^a_n&\mbox{if}~n>m,
\end{array}\right.\nonumber
\end{equation}
and correspondingly for the antiholomorphic sector. The central charge is given by
\begin{equation}
c=\frac{3k}{k-2}.\nonumber
\end{equation}

The space $\mathcal V_{\mbox{\scriptsize H}_3^+}$ decomposes into irreducible representations of the current algebra as
\begin{equation}
\label{decomposition}
\mathcal V_{\mbox{\scriptsize H}_3^+}=\int^{\oplus}_{\mathcal C^+}dj\mathcal R_j,
\end{equation}
where $\mathcal C^+=-1/2+i\mathbb R_+$. The module $\mathcal R_j$ is constructed standardly. As a first step, one considers $\mathcal P_j=P_j\otimes P_j$, where $P_j$ denotes the unitary principal series of the $\mathfrak{sl}(2,\mathbb C)$ algebra  generated by the zero-modes. As pointed out in \cite{Teschner:1997ft}, these principal series can be realized on the Schwartz space of functions on $\mathbb C$ by means of the differential operators 
\begin{equation}
\mathcal D^+_j=x^2\partial_x-2jx, \qquad\qquad \mathcal D^0_j=x\partial_x-j,\qquad\qquad\mathcal D^-_j=\partial_x,\nonumber
\end{equation}
together with their complex conjugates. $\mathcal P_j$ is then extended to a representation of the full current algebra by requiring $J_n^a \mathcal P_j=\bar J_n^a \mathcal P_j=0$ for $n>0$, and freely generating $\mathcal R_j$ by acting with $J_n^a$ and $\bar J_n^a$ for $n<0$.

Representations $P_j$ and $P_{-1-j}$, and thus $\mathcal P_j$ and $\mathcal P_{-1-j}$, are isomorphic, and so are their affine extensions $\mathcal R_j$ and $\mathcal R_{-1-j}$. The form of the unitary intertwining operator will be given explicitly below. It is useful to extend ${\cal V}_{\mbox{\scriptsize H}_3^+}$ by setting $\mathcal C=-1/2+i\mathbb R$ in Eq.~(\ref{decomposition}) instead of $\mathcal C^+$, and then quotienting the model by the aforementioned equivalence.

Every state $|j,x\rangle \in \mathcal V_{\mbox{\scriptsize H}_3^+}$ is associated with a vertex $\Phi_j(x|z)$, $x,z \in \mathbb C$, by virtue of the state-operator correspondence, {\em i.e.}
\begin{equation}
\label{stateoperator}
|j,x\rangle=\lim_{z\rightarrow0}\Phi_j(x|z)|0\rangle,\qquad\qquad\langle j,x|=\lim_{z\rightarrow\infty}|z|^{4\Delta}\langle0|\Phi_{-1-j}(x|z).
\end{equation}
The vertex operators satisfy the following OPE with the currents,
\begin{equation}
J^a(z)\Phi_j(x|w)\sim-\frac{D^a_j\Phi_j(x|w)}{z-w},\nonumber
\end{equation}
and similarly for the antiholomorphic currents. The operator $\Phi_j(x|z)$ is not only an affine primary but also a primary for the Sugawara-Virasoro algebra, its conformal weight being
\begin{equation}
\Delta_0=-b^2j(1+j),\qquad b^2=\frac{1}{k-2}.\nonumber
\end{equation}
Semiclassically, $\Phi_j(x|z)$ can be identified with the wave function
\begin{equation}
\Psi_j(x|z)=\frac{1+2j}{\pi}\left[\left|\gamma(z)-x\right|^2e^{\phi(z)}+e^{-\phi(z)}\right]^{2j},\nonumber
\end{equation}
where $(\phi,\gamma,\bar\gamma)$ are the Poincar\'e coordinates on H$_3^+$. The quantum operator lacks such a simple expression because of normal ordering. Nevertheless, in the large-$\phi$ regime the interaction vanishes and $\Phi_j(x|z)$ acquires the following form,
\begin{equation}
\label{asymptotic}
\Phi_j(x|z)\sim:e^{-2(1+j)\phi(z)}:\delta\left(\gamma(z)-x\right)+B_j:e^{2j\phi(z)}:\left|\gamma(z)-x\right|^{4j},
\end{equation}
with
\begin{equation}
B_j=-\nu(b)^{1+2j}\frac{1+2j}{\pi}\frac{\Gamma(1+b^2(1+2j))}{\Gamma(1-b^2(1+2j))},\qquad\nu(b)=\pi\frac{\Gamma(1-b^2)}{\Gamma(1+b^2)}.\nonumber
\end{equation}

As proved in \cite{Teschner:1997ft}, the asymptotic expression given by Eq.~(\ref{asymptotic}) fixes a normalization of $\Phi_j(x|z)$ consistent with the following two-point function:
\begin{equation}
\label{2pt}
\left\langle\Phi_{j_1}(x_1|z_1)\Phi_{j_2}(x_2|z_2)\right\rangle=\left[\delta\left(j_{12}^+\right)\delta\left(x_{12}\right)+B_{j_1}\delta\left(j_{12}\right)\left|x_{12}\right|^{4j_1}\right]\left|z_{12}\right|^{-4\Delta_{01}},
\end{equation}
where $x_{12}=x_1-x_2$, $z_{12}=z_1-z_2$, $j_{12}=j_1-j_2$, $j_{12}^+=1+j_1+j_2$, and $\Delta_{01}$ stands for the conformal weight associated with $j_1$. The first term of this correlator is a contact term, while the one smeared over the boundary of H$_3^+$ is the so-called bulk or regular term.

As we have already pointed out, representations $\mathcal P_j$ and $\mathcal P_{-1-j}$, and thus $\mathcal R_j$ and $\mathcal R_{-1-j}$, are equivalent. The associated reflection symmetry is explicitly given by
\begin{equation}
\label{reflexion}
\Phi_{j}(x|z)=R_j\left(\mathcal I_j\Phi_{-1-j}\right)(x|z),
\end{equation}
where the reflection amplitude is
\begin{equation}
\label{fifteen}
R_j=\frac{\pi}{1+2j}B_j=\nu(b)^{1+2j}\frac{\gamma(1+b^2(1+2j))}{b^2(1+2j)},
\end{equation}
and the intertwining operator $\mathcal I_j$ is defined by
\begin{equation}
\label{intertwining}
\left(\mathcal I_j\Phi_{-1-j}\right)(x|z)=\frac{1+2j}{\pi}\int_{\mathbb C}d^2x'\left|x-x'\right|^{4j}\Phi_{-1-j}(x'|z).
\end{equation}
The overall factor in Eq.~(\ref{intertwining}) was chosen so that $\mathcal I_j\circ\mathcal I_{-1-j}=$Id, implying, by virtue of Eq.~(\ref{stateoperator}), its unitarity for $j\in -1/2+i\mathbb R$. Notice that a single reflection in the two-point function swaps the contact and the bulk terms, leaving the propagator unchanged, as expected.

Following \cite{Teschner:1997ft}, states $|j,x\rangle \in \mathcal V_{\mbox{\scriptsize H}_3^+}$ and their duals will be referred to as  ``macroscopic states''. They are understood as distributions on dense subspaces of $\mathcal V_{\mbox{\scriptsize H}_3^+}$ and are delta-function normalizable, as it can be read from Eq.~(\ref{2pt}). Analytic continuations in $j$ of these states deserve the name of ``microscopic states'', and their consideration is crucial for the study of the string in AdS$_3$. Moreover, in order to give account of long string configurations in AdS$_3$, we shall need to relax the strong constraint and consider vertex operators built on principal series with possibly different holomorphic and antiholomorphic spins. We will refer to them as microscopic states as well.

\section{The SL$(2,\mathbb R)$-WZNW model and spectral flow}
\label{sect3}

The standard spectrum of the SL$(2,\mathbb R)$-WZNW model is obtained after imposing the Virasoro constraints on the affine extensions of $\mathcal C_{j\alpha}=C_{j\alpha}\otimes C_{j\alpha}$ and $\mathcal D^{\pm}_j=D^{\pm}_j\otimes D^{\pm}_j$, where $C_{j\alpha}$, $\alpha \in [0,1)$, $j\in\mathcal C$, denotes the principal continuous representations and $D^{\pm}_j$, $j<-1/2$, are the principal discrete series of $\mathfrak{sl}(2,\mathbb R)$ generated by the zero-modes $J^a_0$ and $\bar J^a_0$, respectively for the left and the right sectors. It is known from harmonic analysis that these left-right symmetric combinations of unitary irreducible representations form a complete basis in $\mathcal L^2($AdS$_3)$. 

This space of states, if thought of as the complete spectrum, gives rise to two problems. On the one hand, unitarity imposes a seemingly arbitrary upper bound on the mass of string states in AdS$_3$, so that the internal energy of the string could not be too high. On the other hand, it gives no account of long string configurations, known to be present in the theory from both the classical and semiclassical approaches.

Both puzzles were addressed in \cite{Maldacena:2000hw} (see also \cite{Henningson:1991jc}), where the spectrum was proposed to be enhanced by the so-called spectral flow. Spectral flow automorphisms are parametrized by an integer number $\omega$, known as the spectral flow charge, classically related to the number of times a long string winds around the AdS$_3$ boundary. Given such number, the corresponding map is defined by
\begin{equation}
\label{auto1}
    J^3_n \rightarrow J^3_n - \frac{k}{2}\omega \delta_{n,0},
\end{equation}
\begin{equation}
    J^{\pm}_n \rightarrow J^{\pm}_{n\pm \omega},
\end{equation}
for $n\in\mathbb Z$, and similarly for the antiholomorphic currents. This current algebra isomorphism turns to be also a Virasoro-Sugawara automorphism with
\begin{equation}
\label{auto3}
    L_n \rightarrow L_n+\omega J_n^3-\frac{k}{4}\omega^2\delta_{n,0},
\end{equation}
for $n\in\mathbb Z$, and correspondingly for the $\bar L_n$. 

Unlike rational models, which have underlying compact group symmetries, the spectral flow automorphisms generally give rise to non-equivalent representations when acting on a current module. An exception is given by the case of the spectral flow image of a representation induced by a lowest-weight discrete series with spin $j$, which is isomorphic to one built upon a highest-weight discrete series with a ``reflected'' spin $-k/2-j$ and an additional unit of flow. This module isomorphism, referred to as the series identification, restricts the discrete representations allowed in the spectrum to be either those induced by the lowest or by the highest-weight series, while constraining the spin to lie in the real interval\footnote{Standard $\mathcal L^2$ normalization on AdS$_3$ gives rise to the upper bound (see \cite{Maldacena:2000hw}). The lower
bound arises from spectral flow considerations.}
\begin{equation}
-\frac{k-1}{2}<j<-\frac{1}{2}.\nonumber
\end{equation}

It was conjectured in \cite{Maldacena:2000hw} that the full spectrum of the SL$(2,\mathbb R)$-WZNW consists of the standard spectrum as well as their spectral flow images, running over all possible values of $\omega$. Explicitly, the spectrum of the model $\mathcal V_{\mbox{\scriptsize SL}(2,\mathbb R)}$ decomposes as
\begin{eqnarray}
&& \mathcal V_{\mbox{\scriptsize SL}(2,\mathbb R)} = \bigoplus_{\omega\in\mathbb Z}\left[\int_{-\frac{k-1}{2}}^{-\frac{1}{2}} dj \, \mathcal D^{\omega}_j + \int_{\mathcal C} dj \int_0^1 d\alpha \, \mathcal C^{\omega}_{j\alpha}\right],\nonumber
\end{eqnarray}
where $\mathcal D^{\omega}_j$ and $\mathcal C^{\omega}_{j\alpha}$ are the spectral flow images of $\mathcal D^{+}_j$ and $\mathcal C_{j\alpha}$, respectively.

A suitable realization of $\mathcal V_{\mbox{\scriptsize SL}(2,\mathbb R)}$ is obtained by means of vertex operators in the so-called $m$-basis, where the label $m$ is introduced in order to keep track of the eigenvalue of $J_0^3$ in the unflowed frame. This basis is the best suited for a Wick rotation from the $H_3^+$-WZNW model as well as to further include a spectral flow charge while computing correlation functions. We shall denote the affine primary fields realizing the unflowed spectrum by $\Phi_j(m|z)$ and their images under a spectral flow automorphism by $\Phi^{\omega}_j(m|z)$. 
The OPE of these fields with the currents and their conformal weights are given by
\begin{equation}
J^{3}(z)\Phi^{\omega}_j(m|w) \sim \frac{m+k\omega/2}{z-w}\Phi^{\omega}_j(m|w),\nonumber
\end{equation}
\begin{equation}
 J^{\pm}(z)\Phi^{\omega}_j(m|w) \sim \frac{\mp j + m}{(z-w)^{1 \pm \omega}}\Phi^{\omega}_j(m\pm 1|w),\nonumber
\end{equation}
\begin{equation}
\Delta=\Delta_0-\omega m - \frac{k}{4} \omega^2,\nonumber
\end{equation}
and analogously for the antiholomorphic counterparts. Let us stress that spectral flowed primaries $\Phi^{\omega}_j(m|z)$ with non-trivial $\omega$ are not affine primaries. However, after the Sugawara construction in the spectral flowed frame, it can be proved that they are indeed conformal primary fields.

Unflowed vertex operators in the Lorentzian model are related to states in the H$_3^+$-WZNW model through the following Mellin-like transform,
\begin{equation}
\label{mellin}
\Phi_j(m|z) = \int d^2x \, \left|x^{j+m}\right|^2 \Phi_{-1-j}(x|z),
\end{equation}
where the integrated operator corresponds either to a macroscopic state if $\Phi_j(m|z)$ lies in a continuous series or to a microscopic state if $\Phi_j(m|z)$ is associated with a state in a discrete series. Notice that (\ref{mellin}) has a meaning as long as $m-\bar m \in \mathbb Z$, a fact that we shall always assume.

After applying this formula to (\ref{2pt}) we get the following expression for the two-point function involving only unflowed states,
\begin{equation}
\label{2ptsl2}
\left\langle\Phi_{j_1}(m_1|z_1)\Phi_{j_2}(m_2|z_2)\right\rangle=\left[\delta\left(j_{12}^+\right)+Y^{-1-j_1}_{m_1}\delta\left(j_{12}\right)\right]\delta^2(m_1+m_2)\left|z_{12}\right|^{-4\Delta_{01}},
\end{equation}
where
\begin{equation}
 Y^j_{m}=\frac{\pi B_j}{\gamma(-2j)}\frac{\gamma(-j-m)}{\gamma(1+j-m)},
  \qquad \qquad \gamma(x)=\frac{\Gamma(x)}{\Gamma(1-\bar x)},\nonumber
\end{equation} 
and
\begin{equation}
\delta^2(m)=\int_{\mathbb C} d^2x \, \left|x^{m-1}\right|^2= 4\pi^2 \delta(m+\bar m)\delta_{m,\bar m}.\nonumber
\end{equation}
In order to obtain Eq.~(\ref{2ptsl2}) we have repeatedly used the following complex extension of Euler integral:
\begin{equation}
\label{euler}
\int_{\mathbb C} d^2x \left|x^{a-1}(1-x)^{b-1}\right|^2=\pi\frac{\gamma(a)\gamma(b)}{\gamma(a+b)},
\end{equation}
and the identity
\begin{equation}
\gamma(x)\gamma(1-\bar x)=1.\nonumber
\end{equation}

The reflection symmetry also has a counterpart in the Lorentzian model. Indeed, Eq.~(\ref{mellin}) when applied to (\ref{reflexion})-(\ref{intertwining}) gives
\begin{equation}
\label{reflem}
\Phi_j(m|z)=Y^{-1-j}_{m}\Phi_{-1-j}(m|z).
\end{equation}
For $j\in\mathcal C$, this formula defines the intertwining between $\mathcal C_{j\alpha}$ and $\mathcal C_{-1-j,\alpha}$. For a real value of $j$, this expression lacks this interpretation and Eq.~(\ref{reflem}) is just a functional relation between the analytic continuations of vertex operators, {\em i.e.} microscopic states.

Correlation functions in the SL$(2,\mathbb R)$-WZNW model can violate spectral flow number conservation according to certain selection rules (see \cite{Maldacena:2001km} for more details), their computation being more involved than those with trivial total spectral flow charge. However, these selection rules state that the two-point function must necessarily preserve the total spectral flow charge. Since different assignments of spectral flow adding up to the same amount only affect the overall worldsheet dependence of correlators, it follows that
\begin{equation}
\label{2ptsl2bb}
\left\langle\Phi^{\omega_1}_{j_1}(m_1|z_1)\Phi^{\omega_2}_{j_2}(m_2|z_2)\right\rangle=\delta_{\omega_1+\omega_2,0}\delta^2(m_1+m_2)\left[\delta\left(j_{12}^+\right)+Y^{-1-j_1}_{m_1}\delta\left(j_{12}\right)\right]\left|z_{12}^{-2\Delta_1}\right|^2.
\end{equation}

Importantly, the reflection symmetry extends to the spectral flowed sector as
\begin{equation}
\label{reflembb}
\Phi^{\omega}_j(m|z)=Y^{-1-j}_{m}\Phi^{\omega}_{-1-j}(m|z).
\end{equation}
We will make extensive use of this property below. 

\section{Winding modes in the space-time picture}
\label{sect4}

As we have already pointed out, the $m$-basis is convenient for introducing spectral flow charges. However, the space-time picture is the best suited for interpreting any result in the context of the AdS/CFT conjecture since the $x$-basis vertex operators are ingredients for the string theory operators describing states created by sources in the boundary of the target space. Indeed, if $\Theta(z)$ is a spinless worldsheet vertex corresponding to the internal CFT, the operator
\begin{equation}
V_{j}(x)\sim \int_{\mathbb C} d^2z \, \Phi_j(x|z) \Theta(z)\nonumber
\end{equation}
can be realized (as long as the scaling dimension of the full vertex equals one) as describing a string state created by a point-like source located at $x$ on the boundary of AdS$_3$. By means of the AdS/CFT correspondence, it can be identified with a CFT operator inserted at the same point. Scattering amplitudes involving operators in the space-time representation and integrated over the string worldsheet acquire a similar interpretation as correlation functions on the dual two-dimensional CFT.

For unflowed primaries, the definition of the coordinate basis vertex operators comes from the Euclidean model as microscopic states, {\em i.e.} through analytic continuation. The corresponding correlators follow analogously from those of the H$_3^+$-WZNW model. Now, when dealing with spectral flowed primary fields the situation is more complicated since these operators generally lie in representations with energy unbounded from below. A solution for this issue was proposed in \cite{Maldacena:2001km}. An arbitrary lowest-energy state can be seen from a spectral flowed frame with $\omega > 0$ as the lowest-weight state of a certain discrete representation of the global algebra generated by the zero-modes with a spin $J$ being equal to $-m-k\omega/2$. Similarly, if the flow number $\omega$ is negative, the associated spectral flow automorphism maps the same state into the highest-weight state of a discrete representation with $J=m+k\omega/2$. The algebra generated by the $J_0^a$ is identified with the space-time isometry algebra acting on the background and the global $\mathfrak{sl}(2,\mathbb C)$ symmetry algebra of the CFT at the boundary. Therefore, vertex operators in the $x$-basis having flowed primaries and their global descendants as moments were proposed as those being relevant for physical applications.

Note that the eigenvalues of the Cartan generators do not necessarily agree and, therefore, it will also be the case for the global right and left-moving spins, namely, spectral flowed vertex in the $x$-basis are no longer expected to be spinless operators, their space-time planar spin being given by the difference between $J$ and $\bar J$. This number has to be an integer in order for the corresponding correlation functions to be single-valued. On the other hand, since the lowest- and highest-weight states both contribute to the same operator, a flowed vertex in the space-time picture is not labelled by the spectral flow number but, strictly speaking, by its absolute value.
We shall denote the flowed vertex operators as $\Phi^{j\omega}_{J}(x|z)$, where $\omega$ is now the (positive) amount of spectral flow and the superscript $j$ was introduced in order to remind the spin of the unflowed state this vertex is built from. These operators should be understood as microscopic states, as for the Euclidean theory, that are not Virasoro primaries and thus not affine primary fields either, although they are quasi-primary affine vertices. We shall argue later about some conformal symmetries they retain.

The transformation between the space-time picture and the $m$-basis is carried out in analogy with (\ref{mellin}), namely,
\begin{equation}
\label{fourierJM}
 \Phi^{j\omega}_{J}(M|z) = \int_{\mathbb C} d^2x \left|x^{J+M}\right|^2 \Phi^{-1-j,\omega}_{-1-J}(x|z),
\end{equation}
where $M$ is the $J^3_0$ eigenvalue. By means of this map, we have 
\begin{equation}
\label{identif33}
 \Phi^{j\omega}_{J}(\pm J|z) \propto \Phi^{\mp\omega}_{j}(\pm J \pm k\omega/2|z).
\end{equation}
As stressed in \cite{Maldacena:2001km},  by virtue of \eqref{fourierJM} and \eqref{identif33}, a given $x$-basis vertex receives contributions from states in both lowest- and highest-weight modules. Unlike for the H$_3^+$-WZNW model, in the Lorentzian theory it is not possible to univocally associate a single irreducible representation of the current algebra to a vertex operator defined in the space-time picture.

If the normalization of $\Phi^{j\omega}_{J}(x|z)$ is defined so that the the relation in (\ref{identif33}) is actually an identity, this can then be used for determining the regular term of the propagator in the $x$-basis. The dependence of the amplitude on the boundary coordinates is fixed once invariance under the global $\mathfrak{sl}(2,\mathbb C)$ symmetry is imposed, so that
\begin{equation}
\left\langle \Phi^{j_1\omega_1}_{J_1}(x_1|z_1) \Phi^{j_2\omega_2}_{J_2}(x_2|z_2)\right\rangle\propto\delta^2(J_{12})\left|x_{12}^{2J_1}\right|^2,\nonumber
\end{equation}
 for $J_1\sim J_2$. Transforming this expression by means of (\ref{fourierJM}) we obtain
\begin{equation}
\left\langle \Phi^{j_1\omega_1}_{J_1}(M_1|z_1) \Phi^{j_2\omega_2}_{J_2}(M_2|z_2)\right\rangle \propto \pi \delta^2(J_{12}) \delta^2(M_1+M_2) \frac{\gamma(1+J_1-M_1)\gamma(-1-2J_1)}{\gamma(-J_1-M_1)},\nonumber
\end{equation}
where the proportionality constant depends on all the parameters, with the exception of the spin projections. This overall factor can thus be computed by setting $M_1=-J_1$ and $M_2=J_2$, after using (\ref{identif33}), together with (\ref{2ptsl2bb}). We obtain
\begin{eqnarray}
\label{regular}
&& \left\langle \Phi^{j_1\omega_1}_{J_1}(x_1|z_1) \Phi^{j_2\omega_2}_{J_2}(x_2|z_2)\right\rangle=\nonumber \\
&& ~~~~~ ~~~~~ -\frac{|1+2J_1|^2}{\pi^2}\delta_{\omega_1\omega_2}\delta^2(J_{12})\left[\delta\left(j_{12}^+\right)+Y^{j_1\omega_1}_{J_1}\delta\left(j_{12}\right)\right]\left|x_{12}^{2J_1}z_{12}^{-2\Delta_1}\right|^2,
\end{eqnarray}
where
\begin{equation}
Y^{j\omega}_{J}=\frac{\pi B_j}{\gamma(-2j)}\frac{\gamma_{\omega}(-1-j-J)}{\gamma_{\omega}(j-J)}, \qquad \qquad
 \gamma_{\omega}(x)=\gamma(x+k\omega/2),\nonumber
\end{equation}
and
\begin{equation}
\Delta=\Delta_0-\omega-\omega J+\frac{k}{4}\omega^2.\nonumber
\end{equation}
In the unflowed limit, namely, for $\omega_1=0$ and $J_1,\bar J_1\rightarrow j_1$, this expression reduces to the regular term of (\ref{2pt}).

Notice that a possible contact term in the two-point function of operators in the spectral flowed sector cannot be determined using this type of arguments. Indeed, such a term should only be relevant for $J_1\sim -1-J_2$, which would prevent us from setting both spin projections to extremal weights while attempting to use Eq.~(\ref{identif33}) as before.

\section{Reflection symmetry in the winding sector}
\label{sect5}

A natural way to look for the contact term of the two-point function is by making use of the reflection symmetry, since, as we have already mentioned, a reflection operated in one vertex in the propagator would swap its contact and its regular terms. In this section we describe how such an operation can be defined for vertex operators with non-trivial spectral flow and in the space-time picture. 

Eq.~(\ref{reflembb}) constitutes the naive extension of the reflection symmetry to spectral flowed sectors in the $m$-basis. We can read the effect of this symmetry in the $x$-basis by using, again, Eq.~(\ref{identif33}). By virtue of this equation, it follows from (\ref{reflembb}) that
\begin{equation}
\label{refletrucha}
\Phi^{j\omega}_{J}(\pm J|z) = Y^{-1-j,\omega}_{-1-J} \Phi^{-1-j,\omega}_{J}(\pm J|z).
\end{equation}
The states appearing on the left- and right-hand side in this expression and their global descendants contribute to $\Phi^{j\omega}_{J}(x|z)$ and $\Phi^{-1-j,\omega}_{J}(x|z)$, respectively. Moreover, Eq.~(\ref{refletrucha}) remains valid for any weight, as can be easily seen simply by acting with rising and lowering operators on both sides. Thus, by shifting to the $x$-basis we can write\footnote{This relation will be checked explicitly for operators with unit winding in section \ref{sect7}.} 
\begin{equation}
\label{refletrucha2}
\Phi^{j\omega}_{J}(x|z) = Y^{j\omega}_{J} \Phi^{-1-j,\omega}_{J}(x|z).
\end{equation}
Note, however, that this $\mathbb Z_2$-symmetry does not constitute a ``genuine'' reflection symmetry in space-time. In particular, it does not generate any contact term for the propagator when acting on (\ref{regular}), which is actually left invariant. In other words, Eq.~(\ref{refletrucha2}) is merely a remnant of the reflection symmetry in the unflowed frame. 

As opposed to \eqref{refletrucha2}, a well-suited reflection should reduce to an integro-differential expression reducing to (\ref{intertwining}) upon setting $\omega=0$. In the very same way that the intertwining operator adjusts the asymptotic behavior of $\Phi_{-1-j}(x|z)$ to that of $\Phi_{j}(x|z)$ in the H$^+_3$-WZNW model, the symmetry we seek should properly change the dependence of $\Phi^{-1-j,\omega}_{-1-J}(x|z)$ on the worldsheet coordinates to that of $\Phi^{j\omega}_{J}(x|z)$ as well. As for the Euclidean case, we furthermore expect an integration over the worldsheet with a power-law kernel, namely, an identity of the form
\begin{equation}
\label{reflexion222}
\Phi^{j\omega}_{J}(x|z)=R^{j\omega}_J\left(\mathcal I^{j\omega}_J\Phi^{j\omega}_{-1-J}\right)(x|z),
\end{equation}
with
\begin{equation}
\label{refle1}
\left(\mathcal I^{j\omega}_J\Phi^{j\omega}_{-1-J}\right)(x|z)=\frac{1}{\pi^2}|1+2J||1+\alpha|\int_{\mathbb C}d^2x'd^2z'\left|\left(x-x'\right)^{2J}(z-z')^{\alpha}\right|^2\Phi^{-1-j,\omega}_{-1-J}(x'|z'),
\end{equation}
where $\alpha$ could depend, in principle, on $j$, $J$, and $\omega$. As already stated for the unflowed reflection formula for $j \notin - 1/2 + i \mathbb{R}_+ $ below Eq.~(\ref{reflem}), its generalization for the spectrally flowed case must be seen as a functional relation between microscopic vertex operators. Formulas (\ref{reflexion222})-(\ref{refle1}) do not necessarily imply any equivalence between irreducible representations for the SL(2,$\mathbb R$)-WZNW model current algebra.

In order to explicitly determine the quantities $\alpha$ and $R^{j\omega}_J$ appearing in \eqref{reflexion222} and \eqref{refle1}, let us apply these to both vertex operators in a two-point function. More precisely, we ask for 
\begin{eqnarray}
\label{Kyalpha}
&& \left\langle \Phi^{j_1\omega_1}_{J_1}(x_1|z_1) \Phi^{j_2\omega_2}_{J_2}(x_2|z_2)\right\rangle= B^{j_1\omega_1}_{J_1}B^{j_2\omega_2}_{J_2}\int_{\mathbb C}d^2x_1'd^2x_2'd^2z_1'd^2z_2'\left|\left(x_1-x'_1\right)^{2J_1}(z_1-z'_1)^{\alpha_1}\right.\times \nonumber \\
&& ~~~~~ ~~~~~ ~~~~~ ~~~~~ \left.\left(x_2-x'_2\right)^{2J_2}(z_2-z'_2)^{\alpha_2}\right|^2\left\langle \Phi^{-1-j_1,\omega_1}_{-1-J_1}(x'_1|z'_1) \Phi^{-1-j_2,\omega_2}_{-1-J_2}(x'_2|z'_2)\right\rangle 
\end{eqnarray}
to hold, where, in analogy with \eqref{fifteen}, we have introduced
\begin{equation}
B^{j\omega}_J=\frac{1}{\pi^2}|1+2J||1+\alpha|R^{j\omega}_J.
\end{equation}
Under the assumption that $J=J_1\sim J_2$, the correlators on both sides of this expression take the form in Eq.~(\ref{regular}). We thus get
\begin{eqnarray}
\label{Kyalpha2}
&& \left[\delta\left(j_{12}^+\right)+Y^{j_1\omega}_{J}\delta\left(j_{12}\right)\right]\left|x_{12}^{2J}z_{12}^{-2\Delta}\right|^2= B^{j_1\omega}_{J}B^{j_2\omega}_{J}\int_{\mathbb C}d^2x_1'd^2x_2'd^2z_1'd^2z_2'\left|\left(x_1-x'_1\right)^{2J}(z_1-z'_1)^{\alpha_1}\right. \nonumber \\
&& \times \left.\left(x_2-x'_2\right)^{2J}(z_2-z'_2)^{\alpha_2}\right|^2\left[\delta\left(j_{12}^+\right)+Y^{-1-j_1,\omega}_{-1-J}\delta\left(j_{12}\right)\right]\left|x'^{-2-2J}_{12}z'^{-2\Delta-2\omega(1+2J)}_{12}\right|^2,
\end{eqnarray}
where $\omega=\omega_1=\omega_2$. If, in addition, we set $j=j_1\sim j_2$, it follows that
\begin{eqnarray}
&& Y^{j\omega}_{J}\left|x_{12}^{2J}z_{12}^{-2\Delta}\right|^2= \left(B^{j\omega}_{J}\right)^2Y^{-1-j,\omega}_{-1-J}\int_{\mathbb C}d^2x_1'd^2x_2'd^2z_1'd^2z_2'\left|\left(x_1-x'_1\right)^{2J}(z_1-z'_1)^{\alpha}\right|^2\times \nonumber \\
&& ~~~~~ ~~~~~ ~~~~~ ~~~~~ \left|\left(x_2-x'_2\right)^{2J}(z_2-z'_2)^{\alpha}\right|^2\left|x'^{-2-2J}_{12}z'^{-2\Delta-2\omega(1+2J)}_{12}\right|^2.\nonumber
\end{eqnarray}
By using Eq.~(\ref{euler}) in order to compute all integrals above, together with the following identities
\begin{equation}
Y^{j\omega}_J Y^{-1-j,\omega}_J=1, \qquad\qquad Y^{j\omega}_J Y^{j\omega}_{-1-J}= \left[\frac{\pi B_j}{\gamma(-2j)}\right]^2
\frac{\gamma_{\omega}(-1-j-J)\gamma_{\omega}(-j+J)}{\gamma_{\omega}(1+j+J)\gamma_{\omega}(j-J)}\nonumber
\end{equation}
we obtain
\begin{equation}
\alpha = -1+\omega(1+2J),\nonumber
\end{equation}
and
\begin{equation}
\label{K}
B^{j\omega}_{J}=\frac{iB_j|1+2J|}{\pi\gamma(-2j)\gamma(\omega(1+2J))} \sqrt{\frac{\gamma(2\Delta+2\omega(1+2J))\gamma_{\omega}(-1-j-J)\gamma_{\omega}(-j+J)}{\gamma(2\Delta)\gamma_{\omega}(1+j+J)\gamma_{\omega}(j-J)}},
\end{equation}
so that
\begin{equation}
\label{R}
R^{j\omega}_{J}=\frac{i\pi B_j}{\gamma(-2j)}\frac{\gamma(1-\omega(1+2J))}{\omega|1+2J|} \sqrt{\frac{\gamma(2\Delta+2\omega(1+2J))\gamma_{\omega}(-1-j-J)\gamma_{\omega}(-j+J)}{\gamma(2\Delta)\gamma_{\omega}(1+j+J)\gamma_{\omega}(j-J)}}.
\end{equation}
When $\omega=0$ it follows that $\alpha = -1$, trivializing the worldsheet integration in (\ref{refle1}), as expected. Furthermore, by taking $J,\bar J\rightarrow j$ we get
\begin{equation}
B^{j\omega}_{J}\rightarrow\frac{B_j}{V_{\mbox{\scriptsize conf}}},\nonumber
\end{equation}
the conformal volume in this expression cancelling the one coming from the computation of residues in the worldsheet, or, equivalently,
\begin{equation}
R^{j\omega}_{J}\rightarrow R_j.\nonumber
\end{equation}
We thus find a complete agreement of (\ref{refle1}) with (\ref{reflexion})-(\ref{intertwining}) in the unflowed limit.

Setting $j=j_1=j_2$ in (\ref{Kyalpha2}) allowed us to compute both $\alpha$ and $B^{j\omega}_{J}$. However, we still need to check that these expressions are consistent with the terms proportional to $\delta(j_{12}^{+})$ in (\ref{Kyalpha2}). Since $\alpha$ does not depend on $j$, the worldsheet and space-time integrations in (\ref{Kyalpha2}) give the same result if, instead, we set $j=j_1=-1-j_2$. It follows that  (\ref{Kyalpha2}), and thus (\ref{Kyalpha}), are satisfied as long as
\begin{equation}
B^{j\omega}_{J}B^{-1-j,\omega}_{J}=-\frac{|1+2J|^2}{\pi^4} \frac{\gamma(2\Delta+2\omega(1+2J))}{\gamma(\omega(1+2J))^2\gamma(2\Delta)}.\nonumber
\end{equation}
It is easily seen that the expression obtained in \eqref{K} satisfies this identity.

Last but not least, we also need to check for the idempotence of the reflection symmetry defined by (\ref{refle1}). Indeed, after applying (\ref{refle1}) twice, this follows from the expression for the complex delta function,
\begin{equation}
\label{deltacomp}
\delta(x_{12})=-\frac{|\epsilon|^2}{\pi^2} \int_{\mathbb C} d^2y\, \left|(x_1-y)^{-1+\epsilon}(y-x_2)^{-1-\epsilon}\right|^2,
\end{equation}
used on both the worldsheet and the space-time integrations, together with 
\begin{equation}
R^{j\omega}_{J}R^{-1-j,\omega}_{-1-J}=1,\nonumber
\end{equation}
an identity that straightforwardly follows from \eqref{R}.

\section{The contact term and the full propagator}
\label{sect6}

Eq.~(\ref{refle1}) can be used to compute the contact term of the propagator. For this, we take the two-point function and reflect a single vertex operator, leading to
\begin{eqnarray}
&& \left\langle \Phi^{j_1\omega_1}_{J_1}(x_1|z_1) \Phi^{j_2\omega_2}_{J_2}(x_2|z_2)\right\rangle= B^{j_1\omega_1}_{J_1}\int_{\mathbb C}d^2x_1'd^2z_1' \left|\left(x_1-x'_1\right)^{2J_1}(z_1-z'_1)^{\alpha_1}\right|^2 \times \nonumber \\
&& ~~~~~ ~~~~~ ~~~~~ ~~~~~ \left\langle \Phi^{-1-j_1,\omega_1}_{-1-J_1}(x'_1|z'_1)\Phi^{j_2,\omega_2}_{J_2}(x_2|z_2) \right\rangle.\nonumber
\end{eqnarray}
For $J_1\sim -1-J_2$, the two-point function in the integral can be replaced by the bulk term written in (\ref{regular}), so that
\begin{eqnarray}
 \left\langle \Phi^{j_1\omega_1}_{J_1}(x_1|z_1) \Phi^{j_2\omega_2}_{J_2}(x_2|z_2)\right\rangle= -\frac{|1+2J_1|^2}{\pi^2}\delta_{\omega_1\omega_2}\delta^2(J_{12}^+)B^{j_1\omega_1}_{J_1}\left[\delta\left(j_{12}\right)+Y^{-1-j_1,\omega_1}_{-1-J_1}\delta\left(j_{12}^+\right)\right] \times \nonumber \\
 \int_{\mathbb C}d^2x_1'd^2z_1' \left|\left(x_1-x'_1\right)^{2J_1}(z_1-z'_1)^{-1+\omega_1(1+2J_1)}\left(x_1'-x_2\right)^{-2-2J_1}\left(z_1'-z_2\right)^{-2\Delta_1-2\omega_1(1+2J_1)}\right|^2.\nonumber
\end{eqnarray}
After integrating over $x'$ and $z'$ we obtain
\begin{eqnarray}
&& \left\langle \Phi^{j_1\omega_1}_{J_1}(x_1|z_1) \Phi^{j_2\omega_2}_{J_2}(x_2|z_2)\right\rangle= \delta_{\omega_1\omega_2}\delta^2(J_{12}^+)B^{j_1\omega_1}_{J_1}\left[\delta\left(j_{12}\right)+Y^{-1-j_1,\omega_1}_{-1-J_1}\delta\left(j_{12}^+\right)\right] \times \nonumber \\
&& ~~~~~ ~~~~~ ~~~~~ \pi \frac{\gamma(2\Delta_1+\omega_1(1+2J_1))\gamma(\omega_1(1+2J_1))}{\gamma(2\Delta_1+2\omega_1(1+2J_1))}\delta\left(x_{12}\right)\left|z_{12}^{-2\Delta_1-\omega_1(1+2J_1)}\right|^2, \nonumber
\end{eqnarray}
or, more explicitly,
\begin{eqnarray}
\label{contact4}
 \left\langle \Phi^{j_1\omega_1}_{J_1}(x_1|z_1) \Phi^{j_2\omega_2}_{J_2}(x_2|z_2)\right\rangle=\frac{i|1+2J_1|}{\pi} \delta_{\omega_1\omega_2}\delta^2(J_{12}^+)\left[\delta\left(j_{12}^+\right)+Y^{j_1\omega_1}_{-1-J_1}\delta\left(j_{12}\right)\right] \times \nonumber \\
 \sqrt{\frac{\gamma(2\Delta_1+\omega_1(1+2J_1))^2\gamma_{\omega_1}(1+j_1+J_1)\gamma_{\omega_1}(-1-j_1-J_1)}{\gamma(2\Delta_1)\gamma(2\Delta_1+2\omega_1(1+2J_1))\gamma_{\omega_1}(j_1-J_1)\gamma_{\omega_1}(-j_1+J_1)}}\delta\left(x_{12}\right)\left|z_{12}^{-2\Delta_1-\omega_1(1+2J_1)}\right|^2.
\end{eqnarray}

The full expression of the propagator for the SL$(2,\mathbb R)$-WZNW model is obtained by adding (\ref{contact4}) to (\ref{regular}). A well suited parametrization of the two-point function can be obtained by considering the following {\em ansatz}:
\begin{eqnarray}
\label{2ptdef}
 &&\left\langle \Phi^{j_1\omega_1}_{J_1}(x_1|z_1) \Phi^{j_2\omega_2}_{J_2}(x_2|z_2)\right\rangle = S_{J_1}^{j_1\omega_1}S_{J_2}^{j_2\omega_2}\delta_{\omega_1\omega_2}\left[\delta\left(j_{12}^+\right)+L^{j_1\omega_1}_{J_1}\delta\left(j_{12}\right)\right] \times \nonumber \\
 && ~~~~~ ~~~~~ ~~~~~ ~~~~~ ~~~~~ ~~~~~ ~~~~~ \left[\delta^2\left(J_{12}^+\right)\delta\left(x_{12}\right)+M^{j_1\omega_1}_{J_1}\delta^2\left(J_{12}\right)\left|x_{12}^{2J_1}\right|^2 \right]\left|z_{12}^{-\Delta_1-\Delta_2}\right|^2.
\end{eqnarray}
Notice that the factorization of the overall constant is a highly nontrivial proposal that, if fulfilled, would allow us to absorb this factor through a proper redefinition of the spectral flowed vertex fields. Let us show that this is indeed the case.

It can be seen from (\ref{regular}) and (\ref{contact4}) that the identities 
\begin{equation}
    S_{J}^{j\omega}S_{-1-J}^{-1-j,\omega}=\frac{i|1+2J|}{\pi}\sqrt{\frac{\gamma(2\Delta+\omega(1+2J))^2\gamma_{\omega}(1+j+J)\gamma_{\omega}(-1-j-J)}{\gamma(2\Delta)\gamma(2\Delta+2\omega(1+2J))\gamma_{\omega}(j-J)\gamma_{\omega}(-j+J)}},\nonumber
\end{equation}
\begin{equation}
    S_{J}^{j\omega}S_{-1-J}^{j\omega}L_{J}^{j\omega}=\frac{i|1+2J|B_j}{\gamma(-2j)}\sqrt{\frac{\gamma(2\Delta+\omega(1+2J))^2\gamma_{\omega}(-j+J)\gamma_{\omega}(-1-j-J)}{\gamma(2\Delta)\gamma(2\Delta+2\omega(1+2J))\gamma_{\omega}(j-J)\gamma_{\omega}(1+j+J)}},\nonumber
\end{equation}
\begin{equation}
    S_{J}^{j\omega}S_{J}^{-1-j,\omega}M_{J}^{j\omega}=-\frac{|1+2J|^2}{\pi^2},\nonumber
\end{equation}
\begin{equation}
    \left(S_{J}^{j\omega}\right)^2L_{J}^{j\omega}M_{J}^{j\omega}=-\frac{|1+2J|^2}{\pi^2}Y_J^{j,\omega}\nonumber
\end{equation}
must hold. The solution is given by 
\begin{equation}
\label{defS}
    S_{J}^{j\omega}=\sqrt{\frac{\gamma(2+2J)\gamma(2\Delta+\omega(1+2J))\gamma_{\omega}(-1-j-J)}{\pi\gamma(2\Delta)\gamma_{\omega}(j-J)}},
\end{equation}
\begin{equation}
\label{defL}
    L_{J}^{j\omega}=\frac{\pi B_j}{\gamma(-2j)},
\end{equation}
\begin{equation}
\label{defM}
    M_{J}^{j\omega}=\frac{\gamma(-2J)\gamma(2\Delta)}{\pi\gamma(2\Delta+\omega(1+2J))}.
\end{equation}
As a consistency check, note that in the unflowed limit these expressions reduce to
\begin{equation}
    S_{J}^{j\omega}\rightarrow V_{\mbox{\scriptsize conf}}^{-1/2}
    \ \ , \ \ 
    L_{J}^{j\omega}\rightarrow\frac{\pi B_j}{\gamma(-2j)}
    \ \ , \ \
    M_{J}^{j\omega}\rightarrow\frac{\gamma(-2j)}{\pi},\nonumber
\end{equation}
so that the conformal volume factors coming from $S_{J_1}^{j_1\omega_1}$ and $S_{J_2}^{j_2\omega_2}$ cancel the divergence coming from the product of delta functions and the expected propagator (\ref{2pt}) is obtained.

As stated in \cite{Maldacena:2001km}, the terms proportional to $\delta\left( j_{12}^+\right)$ in \eqref{2ptdef} are irrelevant for operators describing short strings. This also holds for the contributions with a factor $\delta^2\left(J_{12}^+\right)$ in situations with non-trivial winding, since $J_{12}^+$ also becomes a strictly positive number. As for the unflowed case, this results in the absence of a contact term.  
For the continuous series this is not longer the case since $m_1$ and $m_2$ are allowed to take any real value. Consequently, the contact term has to be taken into account for long strings configurations.

A relevant aspect to point out concerning the propagator is related to the dependence of its contact term \eqref{contact4} on the worldsheet coordinates, which is, indeed, different from that of the regular term \eqref{regular}. This fact shows that, as advertised above, the spectral flowed vertex operators in the space-time picture are not only not conformal primaries, but not even Virasoro quasi-primary fields. As it can be read off from \eqref{2ptdef}, invariance of the two-point functions under special conformal transformations is manifestly broken in spectrally flowed sectors, although they retain their invariance under translations, rotations and dilations. 

We would like to make a final comment about the spectral flowed two-point function in target space. In order to compute the propagator in space-time, we need to consider \eqref{2ptdef}, modify its dependence on the worldsheet coordinates to give account of the internal CFT, then integrate over $z$ and $\bar z$ and divide it by the volume of the conformal group on the sphere. At the end, this produces an additional factor $V_{\mbox{\scriptsize conf}}^{-1}$. Since none of the delta functions appearing in \eqref{2ptdef} needs to be evaluated for the continuous series, a finite result is achieved in string theory by normalizing the vertex fields as $\Phi^{j\omega}_{J}(x|z)\rightarrow \sqrt{V_{\mbox{\scriptsize conf}}}\, \Phi^{j\omega}_{J}(x|z)$. Note that this normalization differs from the one in \cite{Maldacena:2001km}. For short strings, delta functions involving $J_1$ and $J_2$ must be evaluated producing an extra overall factor $V_{\mbox{\scriptsize conf}}$. Therefore, unlike the case of the long string, we do not have to rescale the operator $\Phi^{j\omega}_{J}(x|z)$.

\section{The singly flowed sector}
\label{sect7}

In \cite{Maldacena:2001km} the authors introduced a definition of vertex operators with one unit of spectral flow based on the fusion of the unflowed state $\Phi_{-1-j}(x|z)$ and the so-called spectral flow operator $\Phi_{-k/2}(x|z)$, with no transformation neither from nor to the $m$-basis. In our notation, this definition reads
\begin{eqnarray}
 &&\Psi^{j}_{J}(x|z)=  \frac{|1+2J|^2}{\pi^2} Y_J^{j,\omega=1} \lim_{\epsilon\rightarrow 0} \left|\epsilon^{1+J-k/2}\right|^2 \int_{\mathbb C} d^2x'd^2y \left|(x-x')^{2J}y^{j-1-J+k/2}\right|^2\times \nonumber\\
 && ~~~~~ ~~~~~ ~~~~~ ~~~~~ ~~~~~ ~~~~~ ~~~~~ ~~~~~ \Phi_{-1-j}(x'+y|z+\epsilon)\Phi_{-k/2}(x'|z).
\label{defw1Malda}
\end{eqnarray}

As a quick consistency check of this equation, let us replace the unflowed vertex in the integrand in \eqref{defw1Malda} by means of its reflection (\ref{reflexion}). We obtain
\begin{eqnarray}
 &&\Psi^{j}_{J}(x|z)= \frac{|1+2J|^2}{\pi^2} Y_J^{j,\omega=1} B_{-1-j} \lim_{\epsilon\rightarrow 0} \left|\epsilon^{1+J-k/2}\right|^2 \int_{\mathbb C} d^2x'd^2x''d^2y \times \nonumber\\
 && ~~~~~ ~~~~~ ~~~~~ \left|(x-x')^{2J}(x'+y-x'')^{-2-2j}y^{j-1-J+k/2}\right|^2\Phi_{j}(x''|z+\epsilon)\Phi_{-k/2}(x'|z).\nonumber
\end{eqnarray}
After the integration over $y$ is performed, and defining $u=(x''-x')$, we get
\begin{eqnarray}
 &&\Psi^{j}_{J}(x|z)= \frac{|1+2J|^2}{\pi^2}Y_{J}^{j,\omega=1}
 \frac{\pi B_{-1-j}}{\gamma(2+2j)} \frac{\gamma_{\omega=1}(j-J)}{\gamma_{\omega=1}(-1-j-J)}\lim_{\epsilon\rightarrow 0} \left|\epsilon^{1+J-k/2}\right|^2  \times \nonumber\\
 && ~~~~~ ~~~~~ ~~~~~  
 \int_{\mathbb C} d^2x'd^2u
 \left|(x-x')^{2J}u^{-2-j-J+k/2}\right|^2\Phi_{j}(x'+u|z+\epsilon)\Phi_{-k/2}(x'|z) \nonumber\\
 && ~~~~~ ~~~~~ = Y^{j\omega=1}_{J} \Psi^{-1-j}_{J}(x|z),\nonumber
\end{eqnarray}
which explicitly shows that a reflection in the unflowed sector does not induce the emergence of a reflection in the spectrally flowed case but simply the identification Eq.~(\ref{refletrucha2}).

Starting from \eqref{defw1Malda} and using the four-point function
\begin{eqnarray}
    \label{4pt}
    && \langle \Phi_{j_1}(x_1'+y_1|z_1+\epsilon_1)\Phi_{-k/2}(x_1'|z_1)
    \Phi_{j_2}(x_2'+y_2|z_2+\epsilon_2)\Phi_{-k/2}(x_2'|z_2)
    \rangle = \nonumber \\
    && \ \ B_{j_1}\delta(j_{12})\left|
    z_{21}^{k/2}\left(z_{21}+\epsilon_{21}\right)^{-2\Delta_{01}} z^{-j_1}\left(1-z\right)^{-j_1}
    x_{21}'^{-k}\left(x_{21}'+y_{21}\right)^{2j_1} \left(z-x\right)^{2j_1}
    \right|^2,  
\end{eqnarray}
where the cross ratios are given by 
\begin{equation}
    x = \frac{y_1 y_2}{x_{21}'(x_{21}'+y_{21})}, \qquad \ z = \frac{\epsilon_1 \epsilon_2}{z_{21}'(z_{21}'+\epsilon_{21})},\nonumber
\end{equation}
the authors re-obtained the bulk term of the propagator \eqref{2ptdef} for the case $\omega=1$. More precisely, this method generates only the part of the bulk term that is proportional to $\delta (j_{12})$. In order to get the term proportional to $\delta(j_{12}^+)$ one has to consider a second solution to the Knizhnik–Zamolodchikov equation and null-state condition associated to the four point function on the left-hand side of \eqref{4pt}, which is of the form $z^{j_1} \delta(x-z)$ \cite{Maldacena:2001km}. Of course, this term had to be there, otherwise \eqref{refletrucha2} would lead to an inconsistency. Alternatively, we could obtain the same result by re-defining
\begin{equation}
    \Psi^{j}_{J}(x|z) \to \frac{1}{\sqrt{2}}\left[ \Psi^{j}_{J}(x|z) + Y_J^{j,\omega=1} \Psi^{-1-j}_{J}(x|z)\right],\nonumber
\end{equation}
which is not unexpected since, as stated above, the identity \eqref{refletrucha2} is a manifestation of the reflection symmetry of the unflowed sector of the theory. 

As described in the previous section, the two-point function \eqref{2ptdef} is not given solely by the bulk contribution. It also includes a contact term that cannot be derived from \eqref{defw1Malda}. In order to address this issue our guide will once again be the reflection symmetry in the spectrally flowed sector. Recall that, for unflowed vertex operators, both terms in \eqref{2pt} were simply exchanged upon reflecting one of the operators by using \eqref{reflexion}, leaving the full correlator unchanged. An analogous statement of course holds for the two-point function in the $\omega=1$ sector, where the reflection is now given by \eqref{refle1}. Based on this property, and inspired by the asymptotic expression in Eq. \eqref{asymptotic}, we introduce a simple {\em ansatz} for completing the definition of $\Phi^{j\omega=1}_{J}(x|z)$ in the space-time picture. Concretely, we propose to include the reflected version of the expression in the right-hand side of \eqref{defw1Malda}, {\em i.e.} we define 
\begin{equation}
    \Phi^{j\omega=1}_{J}(x|z) \equiv \frac{1}{\sqrt{2}} \left[
    \Psi^{j}_{J}(x|z) +  R^{j\omega=1}_J\left(\mathcal I^{j\omega=1}_J\Psi^{j}_{-1-J}\right)(x|z)
    \right].
    \label{newdef}
\end{equation}
 Let us show how the new term looks like. By using \eqref{refle1}, and performing a trivial integration in the $x$-variables, we find
\begin{eqnarray}
&& \left(\mathcal I^{j\omega=1}_J\Psi^{j}_{-1-J}\right)(x|z) = 
\frac{1}{\pi^2} |1+2J|^2 \int_{\mathbb C}d^2 x' d^2 z'\left|\left(x-x'\right)^{2J}(z-z')^{2J}\right|^2 \Psi^{-1-j}_{-1-J}(x'|z') \nonumber \\
 && ~~~~~ ~~  = - \frac{1}{\pi^2} |1+2J|^2 Y_{-1-J}^{-1-j,\omega=1}
 \lim_{\epsilon\rightarrow 0} \left|\epsilon^{-J-k/2}\right|^2
 \int_{\mathbb C} d^2z'd^2y \left|(z-z')^{2J}y^{J-1-j+k/2}\right|^2 \times  \nonumber \\
 && ~~~~~ ~~~~~ ~~~~~ ~~~~~ ~~~~~ ~~~~~ ~~~~~ ~~~~~ ~~~~~ ~~~~~ ~~~~~ \times \Phi_j(x+y|z'+\epsilon) \Phi_{-k/2}(x|z').\nonumber
\end{eqnarray}
Equivalently, we can write
\begin{eqnarray}
&& \left(\mathcal I^{j\omega=1}_J\Psi^{j}_{-1-J}\right)(x|z) = - \frac{1}{\pi^2} |1+2J|^2 Y_{-1-J}^{-1-j,\omega=1}
 \lim_{y\rightarrow 0} \left|y^{J-j-k/2}\right|^2 \times \nonumber \\
 && ~~~~~ ~~~~~ ~~~~~ ~~~~~  \times
 \int_{\mathbb C} d^2z'd^2\epsilon \left|(z-z')^{2J}\epsilon^{-1-J-k/2}\right|^2 
 \Phi_j(x+y|z'+\epsilon) \Phi_{-k/2}(x|z').\nonumber
\end{eqnarray}
The last expression was obtained by using the following identity
\cite{Maldacena:2001km}
\begin{eqnarray}
&&\lim_{\epsilon\rightarrow 0} \left|\epsilon^{m}\right|^2
 \int_{\mathbb C} d^2y \left|y^{-1-j-m}\right|^2 \Phi_j(x+y|z+\epsilon) \Phi_{-k/2}(x|z)=\nonumber \\
 && ~~~~~ ~~~~~ ~~~~~ \lim_{y\rightarrow 0} \left|y^{-j-m}\right|^2 \int_{\mathbb C} d^2\epsilon \left|\epsilon^{m-1}\right|^2 
 \Phi_j(x+y|z+\epsilon) \Phi_{-k/2}(x|z).\nonumber
\end{eqnarray}
We see that the integrated variables have shifted from space-time to worldsheet coordinates in comparison to \eqref{defw1Malda}. The new term in the definition of the vertex is thus fully local in space-time.

Our goal, by following the recipe outlined above, is to prove that the definition \eqref{newdef} allows us to obtain the full two-point function, {\em i.e.} including the contact term. The proof is two-fold: we need to compute the corresponding two-point function by considering separately the direct and cross terms in the product $\Phi^{j_1\omega=1}_{J_1}(x_1|z_1)\Phi^{j_2\omega=1}_{J_2}(x_2|z_2)$. The former give rise to the bulk term, while the latter originate the novel contact term. 

We start with the first cross term, which takes the following form:
\begin{eqnarray}
    && R^{j_2\omega=1}_{J_2} \left\langle \Psi^{j_1}_{J_1}(x_1|z_1)\left(\mathcal I^{j_2\omega=1}_{J_2}\Psi^{j_2}_{-1-J_2}\right)(x_2|z_2) \right\rangle = A \times 
    \lim_{\epsilon_1,\epsilon_2\rightarrow 0} \left|\epsilon_1^{1+J_1-k/2}
    \epsilon_2^{-J_2-k/2}\right|^2 \times \nonumber \\
    && ~~~~~ ~~~
    \int_{\mathbb C} d^2x_1'd^2y_1 d^2z'_2 d^2y_2 \left| \left(x_1 - x_1'\right)^{2J_1} \left(z_2 - z_2'\right)^{2J_2} y_1^{j_1 - J_1 -1 + k/2}  y_2^{J_2 - j_2 - 1 + k/2} \right|^2 \times \nonumber \\
    && ~~~~~ ~~~
    \langle \Phi_{-1-j_1}(x_1'+y_1|z_1+\epsilon_1)\Phi_{-k/2}(x_1'|z_1)
    \Phi_{j_2}(x_2+y_2|z_2'+\epsilon_2)\Phi_{-k/2}(x_2|z_2')
    \rangle, 
    \label{cross1}
\end{eqnarray}
with 
\begin{equation}
    A =  -\frac{|1+2J_1|^2}{\pi^2} Y_{J_1}^{j_1,\omega=1} Y_{-1-J_2}^{-1-j_2,\omega=1} B_{J_2}^{j_2 \omega=1}.
    \label{A1At2}
\end{equation}
Inserting \eqref{4pt} and changing variables to 
\begin{equation}
    w_i = \frac{y_i}{x_{21}'} \ , \ \xi_i = \frac{\epsilon_i}{z_{21}'},\nonumber
\end{equation}
for $i=1,2$, with $x_{21}' = x_2 - x_1'$ and $z_{21}' = z_2' - z_1$, \eqref{cross1} becomes
\begin{eqnarray}
    && R^{j_2\omega=1}_{J_2} \left\langle \Psi^{j_1}_{J_1}(x_1|z_1)\left(\mathcal I^{j_2\omega=1}_{J_2}\Psi^{j_2}_{-1-J_2}\right)(x_2|z_2) \right\rangle = A \, B_{-1-j_1}\delta(j_{12}^+) \times \nonumber \\
    && ~~~~~ \lim_{\xi_1,\xi_2\rightarrow 0} \left|\xi_1^{2+J_1+j_1-k/2}
    \xi_2^{1+j_1-J_2-k/2}\right|^2 
    \int_{\mathbb C} d^2x_1'd^2z'_2 \left|    
    \left(x_1-x_1'\right)^{2J_1}
    \left(z_2-z_2'\right)^{2J_2} \right. \times \nonumber \\
    && ~~~~~ \left. x_{21}'^{\, J_2 - J_1 - 1}
    z_{21}'^{\, \, 1+J_1-J_2-k/2-2\Delta_{01}} \right|^2 \int_{\mathbb C} d^2w_1 d^2w_2 \left| w_1^{j_1 - J_1 - 1 + k/2} 
 w_2^{J_2 - 1 - j_2 + k/2} \right. \times \nonumber \\
    && ~~~~~ \left. \left(w_1 w_2 - \xi_1 \xi_2 (1 + w_2 - w_1)\right)^{-2-2j_1}   
\right|^2.
    \label{cross2}
\end{eqnarray}
The last integral over $w_1$ and $w_2$ can be explicitly computed after a further change of variables:
\begin{equation}
    w_1 = \sqrt{sz} \, t, \qquad w_2 = \sqrt{sz} \, t^{-1}. \nonumber
\end{equation}
Recalling that we are only interested in the small-$z$ limit, it reduces to
\begin{equation}
    \delta^2\left(J_{12}^+\right) \frac{\pi \, \gamma_{\omega=1}(j_1 - J_1)
    \gamma(-1-2j_1)}{\gamma_{\omega=1} (-1-j_1-J_1)} \left(\xi_1 \xi_2
    \right)^{-2-j_1-J_1+k/2}.
    \label{Iwresult}
\end{equation}
This formula is important for several reasons. First, we have obtained the non-trivial condition on the weights, {\em i.e.} $J_2=-1-J_1$, through the delta function $\delta^2\left(J_{12}^+\right)$. Moreover, this same condition implies that the exponents of $\xi_1$ and $\xi_2$ in \eqref{cross2} are equal, and these factors are exactly cancelled by the last factor in \eqref{Iwresult}, trivializing the $\xi_{1,2} \to 0$ limit. 
Furthermore, we see that $x_1'$ integral is also greatly simplified. It takes exactly the form in \eqref{deltacomp} up to the overall constant and this means that what we just computed is actually a contact term. The remaining integration over $z_2'$ results in a factor proportional to 
\begin{equation}
\left|z_{12}^{1-k/2-2\Delta_{01}}\right|^2 = \left|z_{12}^{-\Delta_1 - \Delta_2}\right|^2. \nonumber  
\end{equation}
The dependence on $z_{12}$ thus reproduces that of \eqref{2ptdef}.
Putting everything together we find that \eqref{cross2} becomes
\begin{eqnarray}
    && R^{j_2\omega=1}_{J_2} \left\langle \Psi^{j_1}_{J_1}(x_1|z_1)\left(\mathcal I^{j_2\omega=1}_{J_2}\Psi^{j_2}_{-1-J_2}\right)(x_2|z_2) \right\rangle = \nonumber \\
    && - \frac{\pi^4 A B_{-1-j_1} \gamma (1+2J_1 +2\Delta_1)\gamma_\omega (j_1 - J_1)}{
    |1+2J_1|^2 \gamma(2j_1)\gamma(2J_1)\gamma(2\Delta_1) \gamma_\omega(-1-j_1-J_1)} \delta\left(j_{12}^+\right) \delta^2\left(J_{12}^+\right) 
    \delta\left(x_{12}\right) \left|z_{12}^{ -\Delta_1 - \Delta_2}\right|^2.\nonumber
\end{eqnarray}
By means of \eqref{A1At2} one can show this expression exactly reproduces that of the contact term in \eqref{2ptdef} for $\omega=1$. More precisely, we have only obtained one of the two contributions to the contact term. This was to be expected since the reflection \eqref{refle1} acts on both $j_2$ and $J_2$ simultaneously. In order to get the one proportional to $\delta\left(j_{12}\right)$ it is necessary to redo this calculation by using the contact term of the unflowed four-point function mentioned above. 

The second crossed term in the product $\Phi^{j_1\omega=1}_{J_1}(x_1|z_1)\Phi^{j_2\omega=1}_{J_2}(x_2|z_2)$ is computed analogously and renders the same result. Therefore, it just remains to show that the contribution with two reflected operators also gives the usual bulk term. We do not write any details here and simply state that this is indeed the case. The manipulations needed to carry out this computations are similar to those we just used. 

Thus, we conclude that the definition \eqref{newdef} is consistent with the general two-point function in the spectral flowed sector written in Eq.\eqref{2ptdef} for the particular case of unit winding. Of course, it would be interesting to see whether this simple {\em  ansatz} holds for higher-point functions as well. We leave this computation for future work.  

\section{Final remarks}
\label{disc}

Let us recall what we have done and summarize the main results of this paper. First, we briefly reviewed the relevant aspects of the H$_3^+$-WZNW model: conserved currents, spectrum and vertex operators. In particular, we highlighted some important properties of the latter, namely, the reflection symmetry presented in Eqs. \eqref{reflexion}-\eqref{intertwining}, relating $\Phi_j(x|z)$ and $\Phi_{-1-j}(x|z)$, and also the exact two-point function, given in \eqref{2pt}, where the bulk and contact terms are precisely the reflection the one of the other. Then, we studied the Lorentzian counterpart, that is, the SL$(2,\mathbb R)$-WZNW model, focusing on the appearance of the spectrally flowed operators. At the classical level, some of these states are related to long strings winding $\omega$ times around the AdS$_3$ boundary. As a matter of fact, vertex operators can be thought of more intuitively by shifting to the $m$-basis, where they can be constructed by starting with one of the unflowed affine primaries $\Phi_j$ and acting with the spectral flow automorphism characterized by Eqs. \eqref{auto1}-\eqref{auto3}. In the space-time picture, the resulting flowed vertex was denoted $\Phi_J^{j\omega}(x|z)$, where, besides $\omega$, $J$ (and $\bar{J}$, which is omitted) is the relevant quantum number, {\em i.e.} the eigenvalue of the Cartan generator, while $j$ is only written explicitly as a reminder of how the operator was constructed. 

In the spectral flowed sector of the theory only the regular term of the two-point function had been computed so far. Moreover, and not unrelated to this, no counterpart of the reflection symmetry was known. Indeed, a naive extension was introduced in Eq. \eqref{refletrucha2} and, roughly speaking, it merely states the existence of alternative ways to define operators with the same values of the spectral flow charge and spin. We proposed a full-fledged reflection symmetry for operators with non-trivial spectral flow, namely,
\begin{equation}
\label{reflexion222bis}
\Phi^{j\omega}_{J}(x|z)=R^{j\omega}_J\left(\mathcal I^{j\omega}_J\Phi^{j\omega}_{-1-J}\right)(x|z),
\end{equation}
with
\begin{equation}
\left(\mathcal I^{j\omega}_J\Phi^{j\omega}_{-1-J}\right)(x|z)=\frac{\omega}{\pi^2}|1+2J|^2\int_{\mathbb C}d^2x'd^2z'\left|\left(x-x'\right)^{2J}(z-z')^{-1+\omega(1+2J)}\right|^2\Phi^{-1-j,\omega}_{-1-J}(x'|z'),\nonumber
\end{equation}
the reflection amplitude $R^{j\omega}_J$ being defined in \eqref{R}. This property is consistent with the form of the bulk term in the two-point function. Furthermore, it relates operators with spins $J$ and $-1-J$, so that it allowed us to compute exactly the missing contact term in the propagator. We presented the complete form of the correlator in Eq. \eqref{2ptdef}.

Interestingly enough, the factorization of the global factor in  Eqs.~\eqref{2ptdef} as $S_{J_1}^{j_1\omega_1}S_{J_2}^{j_2\omega_2}$, where  $S_{J}^{j\omega}$ is explicitly given by Eq.~\eqref{defS}, suggests a normalization for the flowed vertex operators more suited than the one set when introducing \eqref{identif33} above. More precisely, by rescaling 
\begin{equation}
   \Phi^{j\omega}_{J}(x|z) \rightarrow S_{J}^{j\omega}\Phi^{j\omega}_{J}(x|z),\nonumber
\end{equation}
we get
\begin{eqnarray}
&& \left\langle \Phi^{j_1\omega_1}_{J_1}(x_1|z_1) \Phi^{j_2\omega_2}_{J_2}(x_2|z_2)\right\rangle = \delta_{\omega_1\omega_2} \left[\delta\left(j_{12}^+\right)+L^{j_1\omega_1}_{J_1}\delta\left(j_{12}\right)\right] \times \nonumber \\
 && \qquad \qquad \qquad \qquad \qquad \qquad \left[\delta^2\left(J_{12}^+\right)\delta\left(x_{12}\right)+M^{j_1\omega_1}_{J_1}\delta^2\left(J_{12}\right)\left|x_{12}^{2J_1}\right|^2 \right] \left|z_{12}^{-\Delta_1-\Delta_2}\right|^2,\nonumber
\end{eqnarray}
where $L^{j\omega}_{J}$ and $M^{j\omega}_{J}$ are given by \eqref{defL} and \eqref{defM}, respectively. Moreover, in this normalization the reflection symmetry is still given by \eqref{reflexion222bis}, albeit with a much simpler coefficient given by 
\begin{equation}
R^{j\omega}_J=\frac{\pi B_j\gamma(-2J)}{\pi\gamma(-2j)}\frac{\gamma(1-\omega(1+2J))}{\omega|1+2J|^2}.\nonumber
\end{equation}
Notice that, unlike in the Euclidean case, none of the factors in the propagator coincide with the reflection coefficient.

There have been interesting recent developments in the study of string propagation in AdS$_3$. Most notably, in \cite{Baggio:2018gct,Dei:2018mfl} integrability techniques were used to solve part of the worldsheet dynamics, while in  \cite{Eberhardt:2019qcl,Eberhardt:2019ywk} a symmetric-product orbifold CFT at large $N$ was proposed as the boundary theory, wherein long-string excitations are related to a specific Liouville factor.  Regarding the construction of \cite{Baggio:2018gct,Dei:2018mfl}, only the short-string sector of the model has been described, up to date, by means of integrability, rendering a comparison with our results futile. Indeed, as stated at the end of Sec.~\ref{sect6}, the contact term in (\ref{2ptdef}) do not appear in this sector of the theory. Concerning the analysis of \cite{Eberhardt:2019ywk} (see also \cite{Hikida:2020kil}), the authors have presented closed expressions for several $x$-basis correlators of spectrally flowed states. However, space-time vertex operators defined in these references are somewhat different from those introduced in \cite{Maldacena:2001km}, which were used throughout this work. Indeed, while vertex operators in the former are built upon a specific lowest-weight state by means of translations in $x$-space, a single vertex in the latter gives rise to both $\mathcal D^+_J$ and $\mathcal D^-_J$, as stated in \cite{Maldacena:2001km} and already stressed below Eq.~\eqref{identif33}. Consequently, an explicit agreement at the level of the two-point function seems difficult to obtain. It would be interesting to understand these issues in more detail in the future.


\section*{Acknowledgments}

The work of S.I. is supported by the National Agency for the Promotion of Science and Technology of Argentina (ANPCyT-FONCyT) Grant PICT-2016-1358 and by CONICET under grant no 22920160100060CO. The work of N.K. is supported by the Leverhulme Trust under grant no RPG-2018-153, and additionally by ANPCyT-FONCyT Grants PICT-2017-1647 and PICT-2015-1525. 


\bibliography{refs}

\providecommand{\href}[2]{#2}\begingroup\raggedright\begin{thebibliography}{10}

\bibitem{Seiberg:1990eb}
N.~Seiberg, \emph{{Notes on quantum Liouville theory and quantum gravity}},
  \href{https://doi.org/10.1143/PTPS.102.319}{\emph{Prog. Theor. Phys. Suppl.}
  {\bfseries 102} (1990) 319}.

\bibitem{Witten:1988hc}
E.~Witten, \emph{{(2+1)-Dimensional Gravity as an Exactly Soluble System}},
  \href{https://doi.org/10.1016/0550-3213(88)90143-5}{\emph{Nucl. Phys.}
  {\bfseries B311} (1988) 46}.

\bibitem{Witten:1991yr}
E.~Witten, \emph{{On string theory and black holes}},
  \href{https://doi.org/10.1103/PhysRevD.44.314}{\emph{Phys. Rev.} {\bfseries
  D44} (1991) 314}.

\bibitem{Mukhi:1993zb}
S.~Mukhi and C.~Vafa, \emph{{Two-dimensional black hole as a topological coset
  model of c = 1 string theory}},
  \href{https://doi.org/10.1016/0550-3213(93)90094-6}{\emph{Nucl. Phys.}
  {\bfseries B407} (1993) 667}
  [\href{https://arxiv.org/abs/hep-th/9301083}{{\ttfamily hep-th/9301083}}].

\bibitem{Ghoshal:1995wm}
D.~Ghoshal and C.~Vafa, \emph{{C = 1 string as the topological theory of the
  conifold}}, \href{https://doi.org/10.1016/0550-3213(95)00408-K}{\emph{Nucl.
  Phys.} {\bfseries B453} (1995) 121}
  [\href{https://arxiv.org/abs/hep-th/9506122}{{\ttfamily hep-th/9506122}}].

\bibitem{Giveon:1999px}
A.~Giveon and D.~Kutasov, \emph{{Little string theory in a double scaling
  limit}}, \href{https://doi.org/10.1088/1126-6708/1999/10/034}{\emph{JHEP}
  {\bfseries 10} (1999) 034}
  [\href{https://arxiv.org/abs/hep-th/9909110}{{\ttfamily hep-th/9909110}}].

\bibitem{Giveon:1999tq}
A.~Giveon and D.~Kutasov, \emph{{Comments on double scaled little string
  theory}}, \href{https://doi.org/10.1088/1126-6708/2000/01/023}{\emph{JHEP}
  {\bfseries 01} (2000) 023}
  [\href{https://arxiv.org/abs/hep-th/9911039}{{\ttfamily hep-th/9911039}}].

\bibitem{Banados:1992wn}
M.~Banados, C.~Teitelboim and J.~Zanelli, \emph{{The Black hole in
  three-dimensional space-time}},
  \href{https://doi.org/10.1103/PhysRevLett.69.1849}{\emph{Phys. Rev. Lett.}
  {\bfseries 69} (1992) 1849}
  [\href{https://arxiv.org/abs/hep-th/9204099}{{\ttfamily hep-th/9204099}}].

\bibitem{Beisert:2010jr}
N.~Beisert et~al., \emph{{Review of AdS/CFT Integrability: An Overview}},
  \href{https://doi.org/10.1007/s11005-011-0529-2}{\emph{Lett. Math. Phys.}
  {\bfseries 99} (2012) 3} [\href{https://arxiv.org/abs/1012.3982}{{\ttfamily
  1012.3982}}].

\bibitem{Zirnbauer:1996zz}
M.~R. Zirnbauer, \emph{{Riemannian symmetric superspaces and their origin in
  random-matrix theory}}, \href{https://doi.org/10.1063/1.531675}{\emph{J.
  Math. Phys.} {\bfseries 37} (1996) 4986}
  [\href{https://arxiv.org/abs/math-ph/9808012}{{\ttfamily math-ph/9808012}}].

\bibitem{Zirnbauer:1999ua}
M.~R. Zirnbauer, \emph{{Conformal field theory of the integer quantum Hall
  plateau transition}},  \href{https://arxiv.org/abs/hep-th/9905054}{{\ttfamily
  hep-th/9905054}}.

\bibitem{Bhaseen:1999nm}
M.~J. Bhaseen, I.~I. Kogan, O.~A. Solovev, N.~Tanigichi and A.~M. Tsvelik,
  \emph{{Towards a field theory of the plateau transitions in the integer
  quantum Hall effect}},
  \href{https://doi.org/10.1016/S0550-3213(00)00276-5}{\emph{Nucl. Phys.}
  {\bfseries B580} (2000) 688}
  [\href{https://arxiv.org/abs/cond-mat/9912060}{{\ttfamily
  cond-mat/9912060}}].

\bibitem{Kogan:1999hz}
I.~I. Kogan and A.~M. Tsvelik, \emph{{Logarithmic operators in the theory of
  plateau transition}}, \href{https://doi.org/10.1142/S0217732300000931,
  10.1016/S0217-7323(00)00093-1}{\emph{Mod. Phys. Lett.} {\bfseries A15} (2000)
  931} [\href{https://arxiv.org/abs/hep-th/9912143}{{\ttfamily
  hep-th/9912143}}].

\bibitem{Maldacena:1997re}
J.~M. Maldacena, \emph{{The Large N limit of superconformal field theories and
  supergravity}}, \href{https://doi.org/10.1023/A:1026654312961,
  10.4310/ATMP.1998.v2.n2.a1}{\emph{Int. J. Theor. Phys.} {\bfseries 38} (1999)
  1113} [\href{https://arxiv.org/abs/hep-th/9711200}{{\ttfamily
  hep-th/9711200}}].

\bibitem{Witten:1998qj}
E.~Witten, \emph{{Anti-de Sitter space and holography}},
  \href{https://doi.org/10.4310/ATMP.1998.v2.n2.a2}{\emph{Adv. Theor. Math.
  Phys.} {\bfseries 2} (1998) 253}
  [\href{https://arxiv.org/abs/hep-th/9802150}{{\ttfamily hep-th/9802150}}].

\bibitem{Gaberdiel:2017oqg}
M.~R. Gaberdiel, R.~Gopakumar and C.~Hull, \emph{{Stringy AdS$_{3}$ from the
  worldsheet}}, \href{https://doi.org/10.1007/JHEP07(2017)090}{\emph{JHEP}
  {\bfseries 07} (2017) 090}
  [\href{https://arxiv.org/abs/1704.08665}{{\ttfamily 1704.08665}}].

\bibitem{Eberhardt:2017pty}
L.~Eberhardt, M.~R. Gaberdiel and W.~Li, \emph{{A holographic dual for string
  theory on $\text{AdS}_3 \times \text{S}^3 \times \text{S}^3 \times
  \text{S}^1$ }}, \href{https://doi.org/10.1007/JHEP08(2017)111}{\emph{JHEP}
  {\bfseries 08} (2017) 111}
  [\href{https://arxiv.org/abs/1707.02705}{{\ttfamily 1707.02705}}].

\bibitem{Eberhardt:2018ouy}
L.~Eberhardt, M.~R. Gaberdiel and R.~Gopakumar, \emph{{The Worldsheet Dual of
  the Symmetric Product CFT}},
  \href{https://doi.org/10.1007/JHEP04(2019)103}{\emph{JHEP} {\bfseries 04}
  (2019) 103} [\href{https://arxiv.org/abs/1812.01007}{{\ttfamily
  1812.01007}}].

\bibitem{Eberhardt:2019niq}
L.~Eberhardt and M.~R. Gaberdiel, \emph{{Strings on $\text{AdS}_3 \times
  \text{S}^3 \times \text{S}^3 \times \text{S}^1$}},
  \href{https://doi.org/10.1007/JHEP06(2019)035}{\emph{JHEP} {\bfseries 06}
  (2019) 035} [\href{https://arxiv.org/abs/1904.01585}{{\ttfamily
  1904.01585}}].

\bibitem{Eberhardt:2019qcl}
L.~Eberhardt and M.~R. Gaberdiel, \emph{{String theory on
  $\boldsymbol{\text{AdS}_{\mathbf{3}}}$ and the symmetric orbifold of
  Liouville theory}},
  \href{https://doi.org/10.1016/j.nuclphysb.2019.114774}{\emph{Nucl. Phys.}
  {\bfseries B948} (2019) 114774}
  [\href{https://arxiv.org/abs/1903.00421}{{\ttfamily 1903.00421}}].

\bibitem{Maldacena:1998bw}
J.~M. Maldacena and A.~Strominger, \emph{{AdS(3) black holes and a stringy
  exclusion principle}},
  \href{https://doi.org/10.1088/1126-6708/1998/12/005}{\emph{JHEP} {\bfseries
  12} (1998) 005} [\href{https://arxiv.org/abs/hep-th/9804085}{{\ttfamily
  hep-th/9804085}}].

\bibitem{Maldacena:1998uz}
J.~M. Maldacena, J.~Michelson and A.~Strominger, \emph{{Anti-de Sitter
  fragmentation}},
  \href{https://doi.org/10.1088/1126-6708/1999/02/011}{\emph{JHEP} {\bfseries
  02} (1999) 011} [\href{https://arxiv.org/abs/hep-th/9812073}{{\ttfamily
  hep-th/9812073}}].

\bibitem{Seiberg:1999xz}
N.~Seiberg and E.~Witten, \emph{{The D1 / D5 system and singular CFT}},
  \href{https://doi.org/10.1088/1126-6708/1999/04/017}{\emph{JHEP} {\bfseries
  04} (1999) 017} [\href{https://arxiv.org/abs/hep-th/9903224}{{\ttfamily
  hep-th/9903224}}].

\bibitem{Maldacena:2000hw}
J.~M. Maldacena and H.~Ooguri, \emph{{Strings in AdS(3) and SL(2,R) WZW model
  1.: The Spectrum}}, \href{https://doi.org/10.1063/1.1377273}{\emph{J. Math.
  Phys.} {\bfseries 42} (2001) 2929}
  [\href{https://arxiv.org/abs/hep-th/0001053}{{\ttfamily hep-th/0001053}}].

\bibitem{Maldacena:2000kv}
J.~M. Maldacena, H.~Ooguri and J.~Son, \emph{{Strings in AdS(3) and the SL(2,R)
  WZW model. Part 2. Euclidean black hole}},
  \href{https://doi.org/10.1063/1.1377039}{\emph{J. Math. Phys.} {\bfseries 42}
  (2001) 2961} [\href{https://arxiv.org/abs/hep-th/0005183}{{\ttfamily
  hep-th/0005183}}].

\bibitem{Maldacena:2001km}
J.~M. Maldacena and H.~Ooguri, \emph{{Strings in AdS(3) and the SL(2,R) WZW
  model. Part 3. Correlation functions}},
  \href{https://doi.org/10.1103/PhysRevD.65.106006}{\emph{Phys. Rev.}
  {\bfseries D65} (2002) 106006}
  [\href{https://arxiv.org/abs/hep-th/0111180}{{\ttfamily hep-th/0111180}}].

\bibitem{Israel:2003ry}
D.~Israel, C.~Kounnas and M.~P. Petropoulos, \emph{{Superstrings on NS5
  backgrounds, deformed AdS(3) and holography}},
  \href{https://doi.org/10.1088/1126-6708/2003/10/028}{\emph{JHEP} {\bfseries
  10} (2003) 028} [\href{https://arxiv.org/abs/hep-th/0306053}{{\ttfamily
  hep-th/0306053}}].

\bibitem{Teschner:1997ft}
J.~Teschner, \emph{{On structure constants and fusion rules in the SL(2,C) /
  SU(2) WZNW model}},
  \href{https://doi.org/10.1016/S0550-3213(99)00072-3}{\emph{Nucl. Phys.}
  {\bfseries B546} (1999) 390}
  [\href{https://arxiv.org/abs/hep-th/9712256}{{\ttfamily hep-th/9712256}}].

\bibitem{Teschner:1999ug}
J.~Teschner, \emph{{Operator product expansion and factorization in the H+(3)
  WZNW model}},
  \href{https://doi.org/10.1016/S0550-3213(99)00785-3}{\emph{Nucl. Phys.}
  {\bfseries B571} (2000) 555}
  [\href{https://arxiv.org/abs/hep-th/9906215}{{\ttfamily hep-th/9906215}}].

\bibitem{Fateev}
V.~Fateev, A.~Zamolodchikov and A.~Zamolodchikov, \emph{{Unpublished notes}}, .

\bibitem{Polchinski:2012nh}
J.~Polchinski and E.~Silverstein, \emph{{Large-density field theory, viscosity,
  and '$2k_F$' singularities from string duals}},
  \href{https://doi.org/10.1088/0264-9381/29/19/194008}{\emph{Class. Quant.
  Grav.} {\bfseries 29} (2012) 194008}
  [\href{https://arxiv.org/abs/1203.1015}{{\ttfamily 1203.1015}}].

\bibitem{Hartnoll:2016apf}
S.~A. Hartnoll, A.~Lucas and S.~Sachdev, \emph{{Holographic quantum matter}},
  \href{https://arxiv.org/abs/1612.07324}{{\ttfamily 1612.07324}}.

\bibitem{Henningson:1991jc}
M.~Henningson, S.~Hwang, P.~Roberts and B.~Sundborg, \emph{{Modular invariance
  of SU(1,1) strings}},
  \href{https://doi.org/10.1016/0370-2693(91)90944-L}{\emph{Phys. Lett.}
  {\bfseries B267} (1991) 350}.

\bibitem{Baggio:2018gct}
M.~Baggio and A.~Sfondrini, \emph{{Strings on NS-NS Backgrounds as Integrable
  Deformations}}, \href{https://doi.org/10.1103/PhysRevD.98.021902}{\emph{Phys.
  Rev. D} {\bfseries 98} (2018) 021902}
  [\href{https://arxiv.org/abs/1804.01998}{{\ttfamily 1804.01998}}].

\bibitem{Dei:2018mfl}
A.~Dei and A.~Sfondrini, \emph{{Integrable spin chain for stringy
  Wess-Zumino-Witten models}},
  \href{https://doi.org/10.1007/JHEP07(2018)109}{\emph{JHEP} {\bfseries 07}
  (2018) 109} [\href{https://arxiv.org/abs/1806.00422}{{\ttfamily
  1806.00422}}].

\bibitem{Eberhardt:2019ywk}
L.~Eberhardt, M.~R. Gaberdiel and R.~Gopakumar, \emph{{Deriving the
  AdS$_{3}$/CFT$_{2}$ corres\-pon\-dence}},
  \href{https://doi.org/10.1007/JHEP02(2020)136}{\emph{JHEP} {\bfseries 02}
  (2020) 136} [\href{https://arxiv.org/abs/1911.00378}{{\ttfamily
  1911.00378}}].

\bibitem{Hikida:2020kil}
Y.~Hikida and T.~Liu, \emph{{Correlation functions of symmetric orbifold from
  AdS$_3$ string theory}},  \href{https://arxiv.org/abs/2005.12511}{{\ttfamily
  2005.12511}}.

\end{thebibliography}\endgroup

\end{document}